**Author's Post-Print (final draft post-refereeing)**

# Snapshot Visualization of Complex Graphs with Force-directed Algorithms

Se-Hang Cheong
Department of Computer and Information Science,
University of Macau
dit.dhc@lostcity-studio.com

Yain-Whar Si
Department of Computer and Information Science,
University of Macau
fstasp@umac.mo

**ABSTRACT**

Force-directed algorithms are widely used for visualizing graphs. However, these algorithms are computationally expensive in producing good quality layouts for complex graphs. The layout quality is largely influenced by execution time and methods' input parameters especially for large complex graphs. The snapshots of visualization generated from these algorithms are useful in presenting the current view or a past state of an information on timeslices. Therefore, researchers often need to make a trade-off between the quality of visualization and the selection of appropriate force-directed algorithms. In this paper, we evaluate the quality of snapshots generated from 7 force-directed algorithms in terms of number of edge crossing and the standard deviations of edge length. Our experimental results showed that KK, FA2 and DH algorithms cannot produce satisfactory visualizations for large graphs within the time limit. KK-MS-DS algorithm can process large and planar graphs but it does not perform well for graphs with low average degrees. KK-MS algorithm produces better visualizations for sparse and non-clustered graphs than KK-MS-DS algorithm.

## 1 INTRODUCTION

In recent years, there has been an intense research activity in graph visualization. A graph is worth a thousand words. Since any relational data could be presented by a graph, it is a popular way for presenting harvested information. Force-directed algorithms are widely used for graph visualization. They can produce visualizations purely based on the structure of a graph and do not require extra attributes. However, there are many types of graphs for different application domains, and each graph has its own unique characteristics such as average degree, density, the distribution of nodes, the distribution of edges, etc. The primary objective of the paper is to review and analyze the performance of available force-directed algorithms for graph visualization. The results obtained from this study can be used for further extending the force-directed algorithms for visualizing large complex graphs.

In section 2, we review the related work. In section 3, we present the experiment settings for analyzing the performance of force-directed algorithms based on different benchmark datasets. In section 4, we compare the algorithms based on a set of attributes. Finally, in section 5, we conclude the paper with future work.

## 2 RELATED WORK

Many force-directed algorithms have been proposed in recent years and there are studies discussed about the performance of these algorithms in graph visualization. Force-directed algorithms rely on spring forces. Forces between the nodes can be computed based on their graph theoretic distances, determined by the lengths of shortest paths between them. Repulsive forces and attractive forces are often used to generate aesthetically pleasing layouts. Graphs drawn with these algorithms tend to exhibit symmetries, and produce crossing-free layouts for planar graphs [1]. Classical force-directed algorithms include Kamada-Kawai (KK) algorithm [2], Davidson Harel (DH) algorithm [3] and Fruchterman Reingold (FR) algorithm [4]. B. Pajntar [5] defined basic properties of the graphs, the criterions of aesthetically pleasing presentation and the characteristics of classical force-directed algorithms. B. Pajntar also summarized aesthetically pleasing presentation of graph visualizations for human readability. Moreover, Brandenburg et al. [6] defined several criteria of aesthetically pleasing visualizations of graphs which include: (a) uniformity of the edge length, (b) uniformity of node distribution, (c) uniformity of edge crossings and (d) display of symmetries. Furthermore, Lipp et al. [7] implemented an extensions of FR algorithm and compare with existing algorithms with respect to the number of edge crossing, standard deviation of edge length and the execution time.

## 3 FORCE-DIRECTED ALGORITHMS

In this section, we introduce seven force-directed algorithms implemented for our experiments: Kamada-Kawai (KK) algorithm [2], Kamada-Kawai with multi-node selection (KK-MS) algorithm [8], Kamada-Kawai with multiple node selection and decaying stiffness (KK-MS-DS) algorithm [8], Davidson Harel (DH) algorithm [3], Fruchterman Reingold (FR) algorithm [4], Fruchterman Reingold algorithm with range extension (FRR) [9] and ForceAtlas2 (FA2) algorithm [10].

### 3.1 Kamada-Kawai (KK) algorithm

The KK algorithm [2] is based on Eades' spring-embedder model [11]. The main objective of the KK algorithm is to distribute the nodes and edges uniformly [12]. To achieve this objective, KK algorithm uses a spring model that minimises the energy function of the graph. The energy function of KK algorithm can be defined as follows:

$$E = \sum_{i=1}^{n-1} \sum_{j=i+1}^{n} \frac{1}{2} k_{i,j} \left( |p_i - p_j| - l_{i,j} \right)^2 \quad (1)$$

where $k_{i,j}$ is the stiffness of the spring between nodes $i$ and $j$, $p_i$ and $p_j$ are the positions of nodes $i$ and $j$ in the visualization and $l_{i,j}$ is the ideal spring distance between nodes $i$ and $j$. The KK algorithm calculates the positions for each pair of nodes, $i$ and $j$. In the visualization, the Euclidean distance of the pair is proportional to $l_{i,j}$.

### 3.2 Kamada-Kawai with multi-node selection (KK-MS) algorithm

The KK algorithm selects and updates a node (with maximum change $\Delta_m$) per iteration. As updating is done one node at a time in the KK algorithm, more iterations are needed when the topology is large, and thus it takes longer to execute. Therefore, KK-MS algorithm inserts the top-$k$ nodes into an ordered queue. Next, KK-MS algorithm pops up the top-$k$ nodes having highest maximum change ($\Delta_m$) from the order queue and updates the nodes' visual positions and corresponding maximum change ($\Delta_m$) values. After that, the algorithm inserts them back into the queue for the next iteration. After $\sqrt{n}$ distinct nodes have been selected for updating their visual positions, KK-MS algorithm clears the ordered queue. When a new ordered queue is created, the KK-MS algorithm recalculates the maximum change ($\Delta_m$) of the nodes from the graph. This process repeats until the termination conditions are met. The time complexity of KK-MS algorithm is $O(n \times (V + v^2))$, where $n$ is the number of iteration, $V$ is the number of nodes in the given graph, and $v$ is the number of nodes in the ordered queue.

### 3.3 Kamada-Kawai with multi-node selection and decaying stiffness (KK-MS-DS) algorithm

The KK algorithm with multi-node selection and decaying stiffness (KK-MS-DS) [8] includes heuristics to achieve faster energy level reduction. The KK-MS-DS algorithm selects nodes with the highest average degrees to be the starting points. Next, the KK-MS-DS algorithm collects all two-hop nodes and constructs an initial starting area. Moreover, KK-MS-DS algorithm adopts the heuristics from the KK-MS algorithm that update a group of $k$ nodes in every iteration, thereby speeding up the updating procedure for the graph. The KK-MS-DS algorithm also uses a decaying stiffness to improve the selection of nodes. That is, the higher the decay rate, the more likely the node is to be selected for the next iteration. In addition, the KK-MS-DS algorithm expands the starting area by checking the stable status ($r$). A stable status implies that a coarse visualization of the starting area has been constructed, but the final stage of the entire graph has not been reached. The ratio of the stable status ($r$) of the starting area is given by:

$$r = \frac{\frac{1}{l}\sum_{i=1}^{l}|\hat{L}_i - L_i|}{\sqrt{\sum_{i=1}^{l}\left(\hat{L}_i - \frac{1}{l}\sum_{i=1}^{l}\hat{L}_i - L_i\right)^2}} \quad (2)$$

where $l$ is the total number of edges in the graph, $\hat{L}_i$ is the edge length of the current iteration, and $L_i$ is the edge length of the input graph. If the stable status is lower than the threshold $\varepsilon$ for a predefined number of iterations, KK-MS-DS algorithm adds outside nodes from the neighbouring area into the starting area.

### 3.4 Davidson Harel (DH) algorithm

The DH algorithm [3] uses a simulated annealing process to produce a visualization in which the nodes are distributed evenly. It is based on the physical annealing process in which liquids are cooled into a crystalline form. This algorithm also prevents nodes from moving too close to non-adjacent edges. An energy value $E$, attraction force $f_a$ and the repulsion force $f_r$ are used in the simulated annealing process. The energy value ($E$) is the sum of all attraction forces and repulsion forces which can be calculated as follows:

$$E = \sum_{i=1}^{n-1}\sum_{j=i+1}^{n} f_a\left(\sqrt{(x_i - x_j)^2 + (y_i - y_j)^2}\right) + f_r\left(\sqrt{(x_i - x_j)^2 + (y_i - y_j)^2}\right) \quad (3)$$

where $i$ and $j$ are nodes and $x_i$ and $y_i$ are coordinates of the node $i$. The attraction force $f_a$ and the repulsion force $f_r$ can be calculated based on equations (6) and (7).

A node $i$ is randomly selected from the graph on initialization. DH algorithm then creates a temporary node $j$, and assigns a position to the node based on the position of node $i$. Therefore, a new energy value $E'$ can be calculated using the position of node $j$ and the other nodes within the graph.

$$E' = \sum_{v,i \in V, j \notin V, v \neq i} f_a\left(\sqrt{(x_v - x_j)^2 + (y_v - y_j)^2}\right) + f_r\left(\sqrt{(x_v - x_j)^2 + (y_v - y_j)^2}\right) \quad (4)$$

where $i$, $j$ and $v$ are nodes and $x_i$ and $y_i$ are coordinates of the node $i$. The attraction force $f_a$ and the repulsion force $f_r$ can be calculated based on equations (6) and (7). If $E' - E \leq 0$ then $E'$ is used as the energy of the next iteration because $E'$ has lower energy value. If $E' - E > 0$, the DH algorithm uses the Boltzmann distribution [13] to determine whether to use the new energy $E'$ in the next iteration. The probability is defined as follows:

$$p = e^{-\frac{(E'-E)}{k \times T}} \quad (5)$$

where $k$ is the Boltzmann constant and $T$ is the temperature variable. If $p$ is less than the threshold $\varphi$, then the new energy $E'$ is accepted; otherwise, the old energy $E$ will be used in the next iteration.

### 3.5 Fruchterman Reingold (FR) Algorithm

The FR [4] algorithm distributes nodes evenly while maintains uniform edge lengths. The FR algorithm uses two forces (attraction and repulsion) to calculate the positions of the nodes, rather than using an energy function with a theoretical graph distance. The attraction force ($f_a$) and repulsion force ($f_r$) are defined as follows:

$$f_a(d) = \frac{d^2}{k} \quad (6)$$

$$f_r(d) = -\frac{k^2}{d} \quad (7)$$

where $d$ is the distance between two nodes and $k$ is the constant of ideal pairwise distance.

### 3.6 Fruchterman Reingold algorithm with range extension (FRR)

FRR algorithm is based on the FR [4] algorithm, but uses a different definition of the ideal pairwise distance $k$. FR algorithm uses an identical ideal pairwise distance $k$ for all edges. FRR algorithm uses the same definitions of the attraction ($f_a$) and repulsion ($f_r$) forces as the FR algorithm. However, FRR algorithm defines $k$ as the distance between corresponding pairs of nodes which can be derived from the weight of links or from the time-of-arrival data [9]. FRR algorithm sometimes fails to generate an acceptable visualization of the given topology as illustrated in Figure 1 (a). If the distribution of the edges is not planar, then many nodes could be stacked together at the centre of the canvas. These distortions are often caused by the attraction force used in the FR algorithm. To alleviate this problem, Efrat et al. [9] proposed a modified version of the attraction force. Using this modified attraction force, nodes can be pulled away from the centre, as illustrated in Figure 1 (b). The enhanced attraction force ($f_a$) is defined as follows:

$$f_a(n_1, n_2) = \frac{d(n_1, n_2)^3}{k} \quad (8)$$

where and $k$ is the ideal pairwise distance and $d$ is the distance between node $n_1$ and node $n_2$

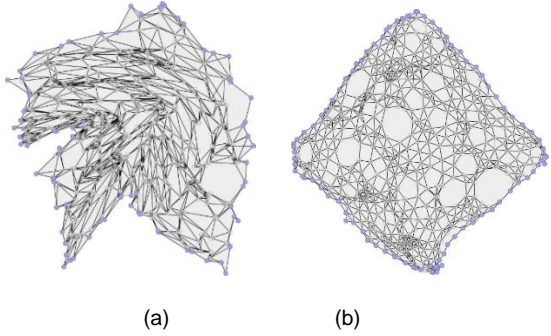

Figure 1 (a) Visualization generated by FRR algorithm, (b) Visualization generated by FRR algorithm with modified attraction force.

### 3.7 ForceAtlas2 (FA2) algorithm

FA2 [10] algorithm which is based on FR algorithm introduces new models of attraction, repulsion, and gravity forces. These models are designed to produces a planer visualization of the graph, yet minimises edge crossing. Firstly, Jacomy et al. [10] used a modified LinLog model as an extension of the attraction model in the FA2 algorithm. The LinLog model was proposed by Andreas Noack [14]. The LinLog model emphasises the visualization of clusters in a graph and tightens those clusters. The modified LinLog model used in the FA2 algorithm is defined as follows:

$$F_a(n_1, n_2) = log(1 + d(n_1, n_2)) \qquad (9)$$

where $d$ is the distance between nodes $n_1$ and $n_2$.

Jacomy et al. [10] proposed a degree-dependent repulsion model for the FA2 algorithm that balance the distance between nodes with higher average degrees and nodes with lower average degrees. This repulsion model increases the chances of nodes with lower average degrees connecting to nodes with higher average degrees. The degree-dependent repulsion model $F_r$ used in the FA2 algorithm is defined as follows:

$$F_r(n_1, n_2) = k \times \frac{(deg(n_1) + 1) \times (deg(n_2) + 1)}{d(n_1, n_2)} \qquad (10)$$

where $d$ is the distance between nodes $n_1$ and $n_2$, $deg(n)$ is the number of edges associated with the node $n$, including in- and out-degree edges and $k$ is a constant of ideal pairwise distance, as used in the FR algorithm. Besides, FA2 algorithm adds the two gravitational forces: gravity and strong gravity. The purpose of the gravity model is to compensate for the repulsion of nodes, and prevents disconnected nodes from drifting away from the centre of the canvas. Jacomy et al. [10] stated that gravity can sometimes be stronger than the attraction and repulsion forces. When the strong gravity model is used in uniform graphs, the nodes can be stacked together. Jacomy et al. [10] also concluded that strong gravity may be useful only for specific types of graphs. The gravity and strong gravity models used in the FA2 algorithm are defined as follows:

$$F_g(n) = k \times (deg(n) + 1) \qquad (11)$$

$$F_{sg}(n) = k \times (deg(n) + 1) \times d(n) \qquad (12)$$

where $k$ is a constant of ideal pairwise distance, $deg(n)$ is the number of edges associated with the node $n$ including in-degree and out-degree edges, and $d(n)$ is the distance from node $n$ to the central point of the canvas.

### 4 EXPERIMENT SETTING

The experiments were performed on an Intel Core i5 CPU with 4 cores, 1.8 GHz and 16 GB RAM running Windows 7. We and implemented all the algorithms in Java and JUNG framework [15]. We implemented seven force-directed algorithms for our experiment. They are Kamada-Kawai (KK) algorithm [2], Kamada-Kawai with multi-node selection (KK-MS) algorithm [8], Kamada-Kawai with multiple node selection and decaying stiffness (KK-MS-DS) algorithm [8], Davidson Harel (DH) algorithm [3], Fruchterman Reingold (FR) algorithm [4], Fruchterman Reingold algorithm with range extension (FRR) [9] and ForceAtlas2 (FA2) algorithm [10].

Our implementations of the algorithms use the same termination criterion. That is, each algorithm terminates when the maximum execution time exceeds 30 minutes. We do not use the maximum iterations for termination criterion because the execution time of an iteration is different in each algorithm. We measured the quality of the visualization (i.e. good visualization) by (a) the number of edge crossings, (b) the standard deviation of the edge lengths. These criteria have been used by other studies to compare the visualization of force-directed algorithms [5-7, 16]. We tested our algorithms on six benchmark data sets which include planar, convex, non-convex, high density degree and non-uniform graphs. The topology information of the data sets is shown in Table 1. These data sets can be downloaded from [17] and they are frequently used in the studies of graph visualization by force-directed algorithms [7, 18-22].

Table 1. Dataset of experiments

| Dataset | Nodes | Edges | Avg. degree |
|---|---|---|---|
| crack | 10240 | 30380 | 5.93 |
| flower_005 | 930 | 13521 | 29.08 |
| grid_rnd_100 | 9497 | 17849 | 3.76 |
| sierpinski_06 | 1095 | 2187 | 3.99 |
| snowflake_B | 971 | 970 | 2 |
| tree_06_03 | 259 | 258 | 1.99 |

### 5 RESULTS

We first compared the number of edge crossings produced by the algorithms. The experimental results are shown in Figure 2. According to results, the data set $crack$ has the largest number of edge crossings and the data set $tree\_06\_03$ has the lowest number of edge crossings among different algorithms. Specifically, the average number of edge crossings in FA2, FRR, DH, FR, KK, KK-DS, KK-MS-DS algorithms are 10803.83, 9694.66, 10754.5, 9603.16, 10227.5, 8378.66, 6337.33 with respect to each data set. From these results, we can observe that KK-MS-DS algorithm achieves the lowest number of edge crossings and FA2 algorithm obtains the highest number of edge crossings.

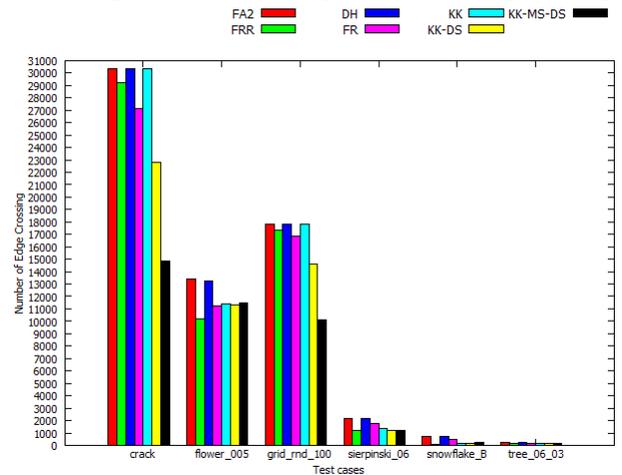

Figure 2 Number of edge crossing for each algorithm

In addition, we compared the standard deviation of edge length in each force-directed algorithm's visualization. The experimental results are illustrated in Figure 3. The standard deviation of edge length of FA2, FRR, DH, FR, KK, KK-DS, KK-MS-DS algorithms are 7.53,

66.69, 104.68, 78.22, 144.56, 12.53, 225.75 with respect to each the data set. We can notice that standard deviation of some the algorithms are quite small. A low standard deviation indicates that the nodes are stacked if the large number of nodes in the snapshot (e.g. FA2 algorithm), while a high standard deviation indicates that the length of edge are spread out (e.g. KK-MS-DS algorithm).

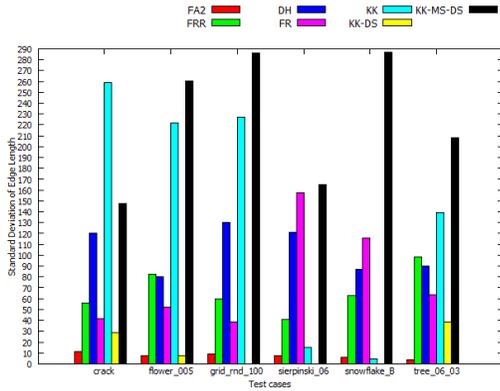

Figure 3 Standard deviation of the edge for each force-directed algorithm

## 5.1 Visualization of force-directed algorithms

In our experiments, the force-directed algorithms are designed to terminate when the execution time exceeds 30 minutes. To better understand the performance of each algorithm, we capture the snapshot of each force-directed algorithm at 10 minutes, 20 minutes and 30 minutes intervals.

From the snapshots, we can observe that although the standard deviation of FA2 algorithm is smaller than other algorithms, the actual visualization does not produce a pleasing graph layout as expected and nodes are stacked each other in the final results. This situation is illustrated in Table 2.

Table 2. The snapshots of visualization by FA2 algorithm.

| Dataset | Snapshot | | |
|---|---|---|---|
| | 10 minutes | 20 minutes | 30 minutes |
| crack | | | |
| flower_005 | | | |
| grid_rnd_100 | | | |
| sierpinski_06 | | | |
| snowflake_B | | | |
| tree_06_03 | | | |

FRR algorithm cannot produces pleasing visualizations for large graphs associated with the data sets $crack$ and $grid\_rnd\_100$ because FRR algorithms need more iterations for large graphs, and thus 30 minutes execution time is still not sufficient under such settings. Besides, the output visualizations for the data sets $tree\_06\_03$ and $sierpinski\_06$ do not expand uniformly and some areas of the graphs are twisted and some of the nodes are stacked. This situation is illustrated in Table 3.

Table 3. The snapshots of visualization by FRR algorithm.

| Dataset | Snapshot | | |
|---|---|---|---|
| | 10 minutes | 20 minutes | 30 minutes |
| crack | | | |
| flower_005 | | | |
| grid_rnd_100 | | | |
| sierpinski_06 | | | |
| snowflake_B | | | |
| tree_06_03 | | | |

DH algorithm is similar to the FRR algorithm (see Table 4). As expected, the algorithm cannot produce pleasing visualizations for large graphs. Besides, DH algorithm does not produce the pleasing visualizations for the datasets except for the $flower\_005$ data set. In addition, the visualizations are neither uniform nor symmetric.

Table 4. The snapshots of visualization by DH algorithm.

| Dataset | Snapshot | | |
|---|---|---|---|
| | 10 minutes | 20 minutes | 30 minutes |
| crack | | | |
| flower_005 | | | |
| grid_rnd_100 | | | |
| sierpinski_06 | | | |

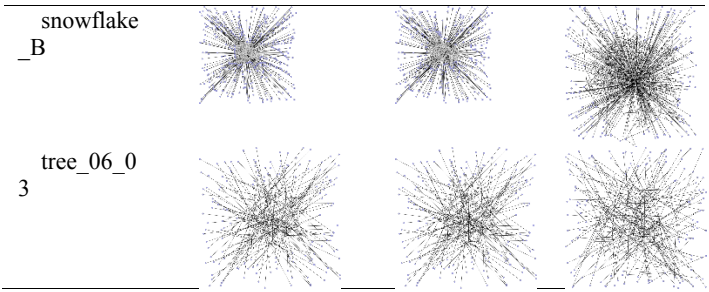

FR algorithm produces better visualizations for data sets *crack* and *grid_rnd*_100. From the snapshots, we can also observe that although some nodes have been expanded by the algorithm, the entire graph is not completely unfolded. However, the visualization of datasets *flower*_005, *sierpinski*_06 and *snowflake_B* are quite poor and FR algorithm could not recover the layout as the FRR algorithm did. This situation is illustrated in Table 5.

Table 5. The snapshots of visualization by FR algorithm.

| Dataset | Snapshot | | |
|---|---|---|---|
| | 10 minutes | 20 minutes | 30 minutes |

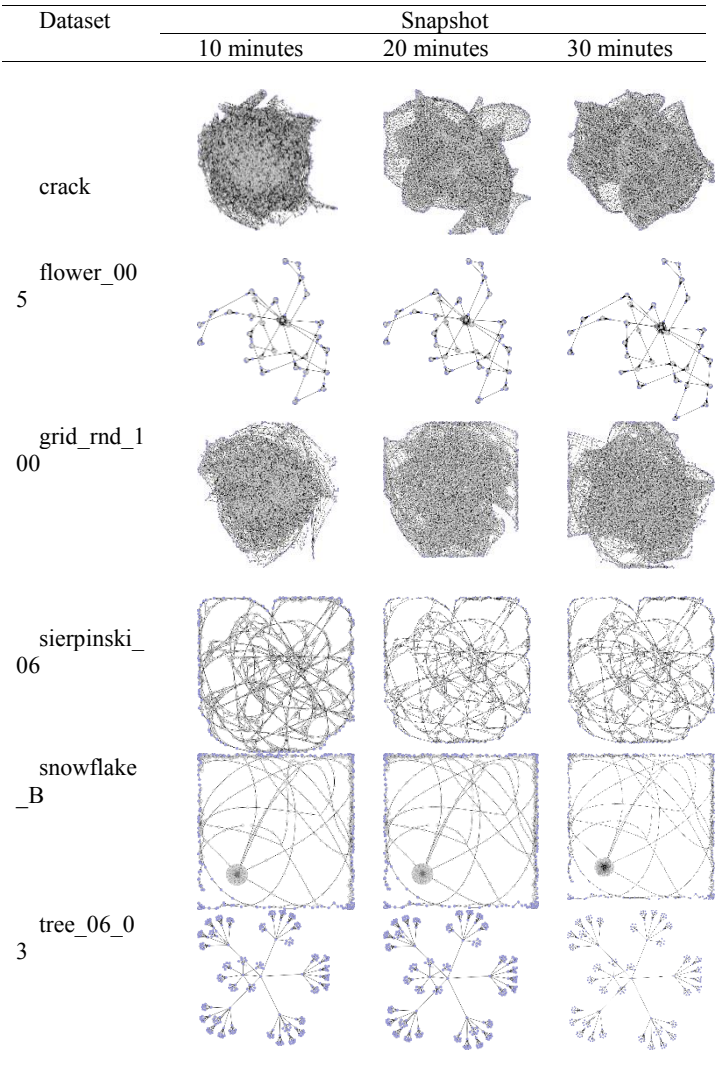

KK algorithm cannot produce pleasing visualizations for large graphs such as data sets *crack* and *grid_rnd*_100 as illustrated in Table 6. Moreover, KK algorithm cannot effectively generate the visualizations for dataset *tree*_06_03. Although the output visualization is symmetric and uniform, the visualization does not look like a tree.

Table 6. The snapshot of visualizations by KK algorithm.

| Dataset | Snapshot |
|---|---|

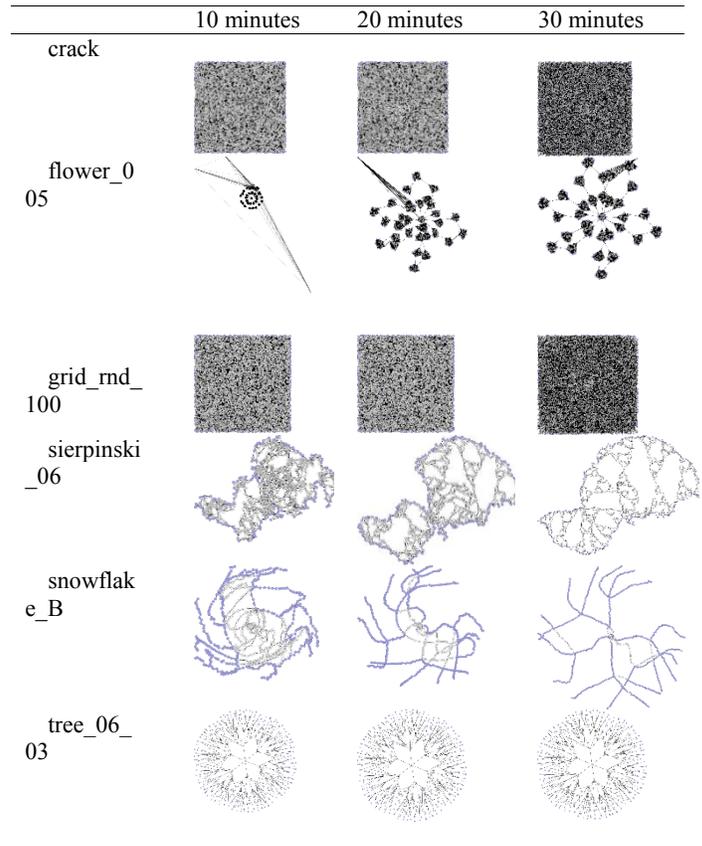

KK-MS algorithm produces pleasing visualizations for datasets *sierpinski*_06 and *snowflake_B*. In addition, some nodes are unfolded in datasets *crack* and *grid_rnd*_100. Similar to the KK algorithm, KK-MS algorithm cannot effectively generate the visualizations for the dataset *tree*_06_03. This situation is illustrated in Table 7.

Table 7. The snapshots of visualization by KK-MS algorithm.

| Dataset | Snapshot | | |
|---|---|---|---|
| | 10 minutes | 20 minutes | 30 minutes |

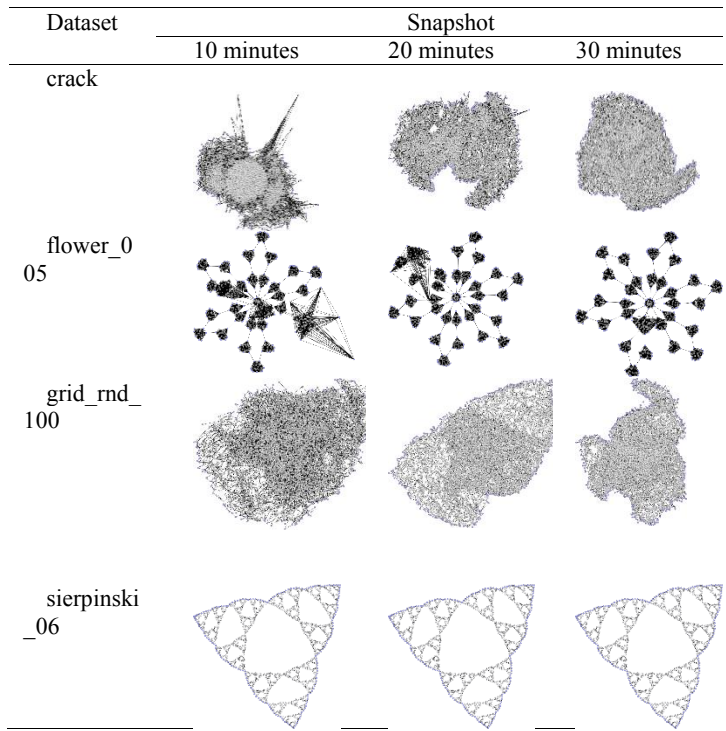

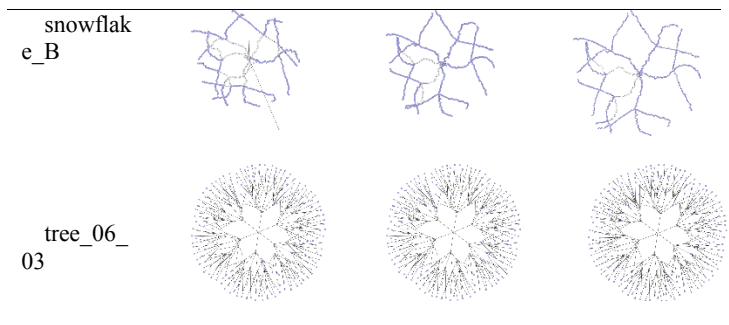

KK-MS-DS algorithm produces pleasing visualizations for datasets *crack* and *grid_rnd*_100 which are planar graphs. In addition, it archives the fastest execution time among all algorithms tested to produce the visualization. However, KK-MS-DS algorithm does not work well for dataset *sierpinski*_06, *flower*_005 and *tree*_06_03. Specifically, KK-MS-DS algorithm does not work well for convex, non-convex and irregular layouts. This situation is illustrated in Table 8.

Table 8. The snapshots of visualization by KK-MS-DS algorithm.

| Dataset | Snapshot | | |
|---|---|---|---|
| | 10 minutes | 20 minutes | 30 minutes |
| crack | | | |
| flower_005 | | | |
| grid_rnd_100 | | | |
| sierpinski_06 | | | |
| snowflake_B | | | |
| tree_06_03 | | | |

## 6 Conclusions

In this paper, we describe a comprehensive performance comparison of seven force-directed algorithms. These algorithms are used to generate snapshot visualizations for a set of complex graphs (i.e. planar graph, non-convex, convex, irregular shapes, etc.). The performance of each algorithm is measured based on the number of edge crossing and the standard deviations of edge length.

From the experiment results, we can observe that KK-MS-DS algorithm performed well for large and planar graphs. It is not only fast but also able to produce pleasing visualizations. However, it does not perform well for sparse and low average degree graphs from data sets $tree\_06\_03$, $sierpinski\_06$ and $snowflake\_B$. KK-MS algorithm however produces better visualizations for sparse and non-clustered graphs, such as $sierpinski\_06$ and $snowflake\_B$. We can observe that the classical KK algorithm is slow. Overall, KK algorithm and its extensions do not perform well for graphs such as $flower\_005$ and $tree\_06\_03$ which contain clusters. FRR and FR algorithms can produce better visualization than KK algorithm for sparse and graphs containing clusters. This situation is captured in visualization for the data sets $flower\_005$ and $tree\_06\_03$. Besides, FR and DH algorithms do not perform well for large and dense graphs.

The experimental results presented in this paper can be used by analysts in selecting suitable force-directed algorithms for visualizing large complex graphs. Although each of the tested algorithms is designed for graph visualizations, there is no single winner when they are applied to complex graphs. We can also notice that a lower standard deviation of edge lengths does not necessarily lead to an aesthetically pleasing graph visualization.

As for the future work, we anticipate the need for faster force-directed approaches for visualizing large complex graphs since the forms of online social interaction continue to evolve and diversify. Therefore, any enabling information visualization technologies such as force-directed algorithms are expected to take on an increasingly important role. At present, we are extending force-directed approaches to visualize large and complex social networks. The extended algorithms are expected to further reduce the execution time by exploiting additional information (attributes) associated with large complex social networks. We are also planning to conduct more rigorous experiments to evaluate the strength and weakness of each algorithm. We are also in the progress of extending current methods to combine with pre-positioning models and multilevel approaches to allow a faster convergence in energy calculation.

### Acknowledgments

This research was funded by the Research Committee of University of Macau, grants MYRG2017-00029-FST and MYRG2016-00148-FST.

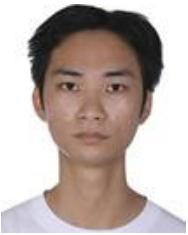
Se-Hang Cheong is a PhD student from the Department of Computer and Information Science at the University of Macau. His current scientific interests are in the areas of graph visualization and wireless sensor network.

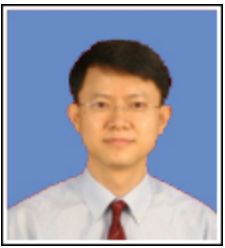
Yain-Whar Si is an associate professor from the Department of Computer and Information Science at the University of Macau. His research interests are in the areas of Data Analytics, Financial Technology (FinTech), Computational Intelligence, Information Visualization, and Business Process Management.